\documentclass[epj]{svjour}
\usepackage{graphics}
\usepackage{amsmath,amssymb}

\begin{document}



\title{Is the exotic $X(5568)$ a bound state?}

\author{Xiaoyun Chen and Jialun Ping
\thanks{Corresponding author: jlping@njnu.edu.cn (J. L. Ping)}}

\institute{Department of Physics and Jiangsu Key Laboratory for Numerical
Simulation of Large Scale Complex Systems, Nanjing Normal University, Nanjing 210023, P. R. China}

\abstract{
Stimulated by the recent observation of the exotic $X(5568)$ state by D0 Collaboration, we 
study the four-quark system $us\bar{b}\bar{d}$ with quantum numbers $J^P=0^+$ in the framework of 
chiral quark model. Two structures, diquark-antidiquark and meson-meson, with all possible 
color configurations are investigated by using Gaussian expansion method. The results show
that energies of the tetraquark states with diquark-antiquark structure are too high to
the candidate of $X(5568)$, and no molecular structure can be formed in our calculations. 
The calculation is also extended to the four-quark system $us\bar{c}\bar{d}$ and the same 
results as that of $us\bar{b}\bar{d}$ are obtained. 
}

\maketitle

\section{Introduction} \label{introduction}

Since the charmonium-like resonance $X(3872)$ is observed by Bell collaboration \cite{x3872} in 2003, a lot of
experiments spring up to study the exotic states-$XYZ$ particles from Belle, BaBar, BESIII, LHCb, CDF, D0 and
other collaborations. And people believe that the traditional convention, the meson is made up of quark and
antiquark as well as baryon is made up of three quarks, is broken. The exotic states were observed in
$B$ meson decays, in $e^{+}e^-$ and $\bar{p}p$ annihilations. In study of $B$ decays, the phenomenon of
$CP$ violation has been studied by experimental collaborations. Many predictions of Standard Model are
confirmed and some hints beyond Standard Model are exposed.

Very recently, the D0 Collaboration observed a narrow structure, named $X(5568)$, in the $B^{0}_s\pi^{\pm}$
invariant mass spectrum with $5.1\sigma$ significance \cite{x5568}. The mass and width measured is
$M=5567.8\pm2.9^{+0.9}_{-1.9}$ MeV and $\Gamma=21.9\pm6.4^{+5.0}_{-2.5}$ MeV, respectively. Its decay mode
$B^{0}_s\pi^{\pm}$ indicates that $X(5568)$ is consist of four different flavors: $u,d,s,b$. $X(5568)$ must
be a $su\bar{b}\bar{d}$ or $sd\bar{b}\bar{u}$ tetraquark state. The D0 Collaboration suggests that the
quantum numbers of $X(5568)$ may be $J^P=0^+$ because $B^{0}_s\pi^{\pm}$ is produced in $S$-wave. However,
the preliminary results of the experimental search of the state by LHCb collaboration is negative \cite{LHCb}.

The discovery of the exotic state $X(5568)$ stimulated the theoretical interest. Many theoretical work has
been done, such as approaches based on QCD sum rules \cite{sumrule1,sumrule2,sumrule3,sumrule4,sumrule5,sumrule6},
quark models \cite{quarkmodel1,quarkmodel2,quarkmodel3}, rescattering effects \cite{scattering}, etc.
Agaev {\em et al.} studied the state $X(5568)$ within the two-point sum rule method using the diquark-antidiquark
interpolating current \cite{sumrule1,ssagaev2} and meson molecule structure \cite{molecule}, their results preferred
diquark-antidiquark picture rather than molecule and a nice agreement with experimental data is obtained.
QCD sum rule method was also employed by other groups to investigate the state $X(5568)$ as the diquark-antidiquark
type scalar and axial-vector tetraquark states \cite{sumrule2,sumrule3,sumrule4,sumrule5,sumrule6}.
In Ref.\cite{quarkmodel1}, a tetraquark interpretation of the $X(5568)$ was proposed based on the diquark-antidiquark
scheme, the identification is possible when the systematic errors of the model is taken into account. This result is
supported by simple quark model estimations \cite{quarkmodel3,quarkmodel4}. The hadronic molecule scenarios of
the $X(5568)$ is also possible according to the calculation of Ref.\cite{quarkmodel2}.
However, there are several theoretical calculations with negative results. Burns and Swanson examined the various
interpretations of the state $X(5568)$ and concluded that the threshold, cusp, molecular and tetraquark models are
all unfavored~\cite{Burns}. F. K. Guo {\em et al.} provided additional arguments using general properties of QCD and obtained
the same conclusion~\cite{FKGuo}. Although the state $X(5568)$ can be reproduced in the coupled channel analysis in
Ref.\cite{XLiu}, the momentum cutoff used is much larger than the normal one.

Considering the quantum numbers $J^P=0^+$ of the state $X(5568)$, the spin and orbit angular momentum can be both taken as zero.
For meson molecule structure, the possible channels are $B^{0}_s\pi$, $B^{*}_s\rho$, $B^+\bar{K}^0$ and $B^{*+}\bar{K^*}^0$.
For diquark-antidiquark structure, the only possible state is $su\bar{b}\bar{d}$ for $X(5568)^+$ or $sd\bar{b}\bar{u}$ for
$X(5568)^-$. In the present work, we compute all these states including molecule and diquark-antidiquark structures using
chiral quark model under the flavor $SU(3)$ and $SU(4)$ symmetry, respectively. Besides, we extend our investigation to
the new family of the four flavor exotic states $X_c$ with $u,d,s,c$ by replacing the $b$ quark with $c$ quark. We hope
that we can find some useful and meaningful information of $X(5568)$ through our systematic calculations.

This article is organized as follows. In Section 2, we introduce the Gaussian Expansion Method(GEM) and chair quark model.
In the next section, the numerical calculations with discussions are shown. A short summary is given in the last section.

\section{GEM and Chiral Quark Model} \label{GEM and chiral quark model}

In the chiral quark model, the mass of the tetraquark state is obtained by solving the Schr\"{o}dinger equation
\begin{equation}
    H \Psi^{IJ}_{M_IM_J}=E^{IJ} \Psi^{IJ}_{M_IM_J}
\end{equation}
where $\Psi^{IJ}_{M_IM_J}$ is the wave function of the tetraquark state, which can be constructed as follows.
First, we write down the wave functions of two clusters (Taking meson-meson configuration as an example),
\begin{eqnarray}
    \Psi^{I_1J_1}_{M_{I_1}M_{J_1}}(12) =\left[ \psi_{l_1}(\mathbf{r}_{12})\chi_{s_1}(12)\right]^{J_1}_{M_{J_1}}
     \omega^{c_1}(12)\phi^{I_1}_{M_{I_1}}(12),  \nonumber \\
    \Psi^{I_2J_2}_{M_{I_2}M_{J_2}}(34) =\left[ \psi_{l_2}(\mathbf{r}_{34})\chi_{s_2}(34)\right]^{J_2}_{M_{J_2}}
    \omega^{c_2}(34)\phi^{I_2}_{M_{I_2}}(34),
\end{eqnarray}
where $\chi_{s},\omega^c,\phi^{I}$ are spin, color and flavor wavefunctions of the
quark-antiquark cluster (the quarks are numbered as 1, 3 and antiquarks 2, 4, see Fig.1).
[~] denotes the angular momentum coupling.
Then the total wave function of tetraquark state is obtained,
\begin{figure}
\resizebox{0.50\textwidth}{!}{\includegraphics{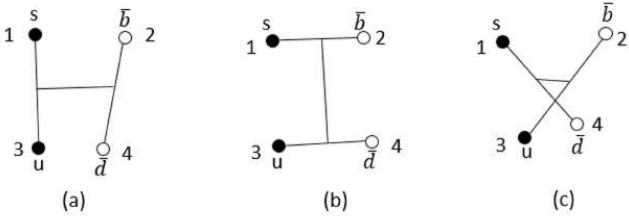}}

\caption{Structure of the tetraquark $us\bar{d}\bar{b}$ system. Solid and open circles represent quarks
and antiquarks, respectively. (a) diquark-antidiquark channel,
(b) direct meson-meson channel: $B_s^0 \pi^+$ or $B_s^* \rho$,
(c) exchange meson-meson channel: $B^+ \bar{K}^0$ or $B^{*+} \bar{K^*}^0$.}
\label{structure}
\end{figure}

\begin{eqnarray}
  \Psi^{IJ}_{M_IM_J} & = & {\cal A} \left[ \Psi^{I_1J_1}(1,2)\Psi^{I_2J_2}(3,4)
     \psi_{L_r}(\mathbf{r}_{1234})\right]^{IJ}_{M_IM_J}    \nonumber \\
  & = & \left[\left[\psi_{l_1}(\mathbf{r}_{12})\chi_{s}(12)\right]^{J_1}
              \left[\psi_{l_2}(\mathbf{r}_{34})\chi_{s}(34)\right]^{J_2} \right. \nonumber \\
  &   & \left. \psi_{L_r}(\mathbf{r}_{1234})\right]^{J}_{M_J} 
        \left[\omega^{c_1}(12)\omega^{c_2}(34)\right]^{[222]} \nonumber \\
  &   & \left[\phi^{I_1}_{M_{I_1}}(12)\phi^{I_2}_{M_{I_2}}(34)\right]^{I}_{M_I},
\end{eqnarray}
where $\psi_{L_r}(\mathbf{r}_{1234})$ is the relative wave function between two clusters
with the relative orbit angular momentum $L_r$. ${\cal A}$ is the antisymmetrization operator.
If all quarks (antiquarks) are taken as identical particles, we have
\begin{equation}
{\cal A}=\frac{1}{2}(1-P_{13}-P_{24}+P_{13}P_{24}).
\end{equation}
In GEM, the orbital wave function is written as the product of radial one and spherical harmonics,
and the radial part of the wavefunction is expanded by gaussians,
\begin{eqnarray}
   \psi_{lm}(\mathbf{r}) &= &\sum_{n=1}^{n_{max}} c_{n}\psi^G_{nlm}(\mathbf{r}), \nonumber \\
   \psi^G_{nlm}(\mathbf{r})& =& N_{nl}r^{l} e^{-\nu_{n}r^2}Y_l^m(\hat{\mathbf{r}}).
\end{eqnarray}
Gaussian size parameters are taken as the following geometric progression numbers
\begin{eqnarray}
    \nu_{n}=\frac{1}{r^2_n}, ~~~~ r_n=r_1a^{n-1},~~~~
    a=\left(\frac{r_{n_{max}}}{r_1}\right)^{\frac{1}{n_{max}-1}}.
\end{eqnarray}
Noting that the gaussians are not orthogonal, the Rayleigh-Ritz variational principle
for solving the Schr\"{o}dinger equation leads to a generalized eigenvalue problem
\begin{eqnarray}
     && \sum_{n^{\prime}~\alpha^{\prime}}(H_{n\alpha,n^{\prime}\alpha^{\prime}}^{IJ}
    -E^{IJ} N_{n\alpha,n^{\prime}\alpha^{\prime}}^{IJ}) C_{n^{\prime}\alpha^{\prime}}^{IJ} = 0, \\
    & & H_{n\alpha,n^{\prime}\alpha^{\prime}}^{IJ} =
      \langle\Phi^{IM_IJM_J}_{n\alpha}| H | \Phi^{IM_IJM_J}_{n^{\prime}\alpha^{\prime}}\rangle ,
      \nonumber \\
    & & N_{n\alpha,n^{\prime}\alpha^{\prime}}^{IJ} =
      \langle\Phi^{IM_IJM_J}_{n\alpha}| \Phi^{IM_IJM_J}_{n^{\prime}\alpha^{\prime}}\rangle,
\end{eqnarray}
where $\alpha$ denotes channels.

The hamiltonian of the chiral quark model includes three parts,
the rest masses of quarks, the kinetic energy and the potential energy. The potential
energy is composed of color confinement, one-gluon-exchange and one Goldstone boson exchange.
The detailed form for tetraquark states is shown below~\cite{chqm}
\begin{eqnarray}
H &=& \sum_{i=1}^4 m_i+\frac{p_{12}^2}{2\mu_{12}}+\frac{p_{34}^2}{2\mu_{34}} 
  +\frac{p_{1234}^2}{2\mu_{1234}}  \nonumber \\
  & &　+\sum_{i<j=1}^4 \left( V_{ij}^{G}+V_{ij}^{C}+\sum_{\chi=\pi,K,\eta} V_{ij}^{\chi}
   +V_{ij}^{\sigma}\right), \nonumber\\
 V_{ij}^{G}&=& \frac{\alpha_s}{4} \boldsymbol{\lambda}_i^c \cdot \boldsymbol{\lambda}_{j}^c
\left[\frac{1}{r_{ij}}-\frac{2\pi}{3m_im_j}\boldsymbol{\sigma}_i\cdot \boldsymbol{\sigma}_j
  \delta(\boldsymbol{r}_{ij})\right],　\nonumber \\
  & &  \delta{(\boldsymbol{r}_{ij})}=\frac{e^{-r_{ij}/r_0(\mu_{ij})}}{4\pi r_{ij}r_0^2(\mu_{ij})},
  \nonumber \\
V_{ij}^{C}&=& ( -a_c r_{ij}^2-\Delta ) \boldsymbol{\lambda}_i^c \cdot
 \boldsymbol{\lambda}_j^c \nonumber \\
V_{ij}^{\pi}&=& \frac{g_{ch}^2}{4\pi}\frac{m_{\pi}^2}{12m_im_j}
  \frac{\Lambda_{\pi}^2}{\Lambda_{\pi}^2-m_{\pi}^2}m_\pi v_{ij}^{\pi}
  \sum_{a=1}^3 \lambda_i^a \lambda_j^a,  \nonumber \\
V_{ij}^{K}&=& \frac{g_{ch}^2}{4\pi}\frac{m_{K}^2}{12m_im_j}
  \frac{\Lambda_K^2}{\Lambda_K^2-m_{K}^2}m_K v_{ij}^{K}
  \sum_{a=4}^7 \lambda_i^a \lambda_j^a,  \nonumber \\
V_{ij}^{\eta} & = & \frac{g_{ch}^2}{4\pi}\frac{m_{\eta}^2}{12m_im_j}
\frac{\Lambda_{\eta}^2}{\Lambda_{\eta}^2-m_{\eta}^2}m_{\eta} v_{ij}^{\eta}
 \left[\lambda_i^8 \lambda_j^8 \cos\theta_P
 - \lambda_i^0 \lambda_j^0 \sin \theta_P \right],  \nonumber \\
V_{ij}^{\sigma}&=& -\frac{g_{ch}^2}{4\pi}
\frac{\Lambda_{\sigma}^2}{\Lambda_{\sigma}^2-m_{\sigma}^2}m_\sigma \left[
 Y(m_\sigma r_{ij})-\frac{\Lambda_{\sigma}}{m_\sigma}Y(\Lambda_{\sigma} r_{ij})\right] \nonumber \\
 v_{ij}^{\chi} & = & \left[ Y(m_\chi r_{ij})-
\frac{\Lambda_{\chi}^3}{m_{\chi}^3}Y(\Lambda_{\chi} r_{ij}) \right]
\boldsymbol{\sigma}_i \cdot\boldsymbol{\sigma}_j, \nonumber \\
& & Y(x)  =   e^{-x}/x,
\end{eqnarray}
where $m_i$ is the mass of quarks and antiquarks, and $\mu_{ij}$ is their reduced mass,
$r_0(\mu_{ij}) =\hat{r}_0/\mu_{ij}$, $\boldsymbol{\sigma}$ are the $SU(2)$ Pauli matrices,
$\boldsymbol{\lambda},~\boldsymbol{\lambda}^c$
are $SU(3)$ flavor, color Gell-Mann matrices, $g^2_{ch}/4\pi$ is the chiral coupling constant,
determined from $\pi$-nucleon coupling constant. $\alpha_s$ is the effective scale-dependent
running quark-gluon coupling constant~\cite{chqm},
\begin{equation}
\alpha_s(\mu_{ij})=\frac{\alpha_0}{\ln\left[(\mu_{ij}^2+\mu_0^2)/\Lambda_0^2\right]}
\end{equation}
All model parameters are determined by fitting the meson spectrum and
shown in Table \ref{modelparameters}. The calculated masses of the mesons involved in
the present work are shown in Table \ref{meson spectrum}.
\begin{table}[htb]
\begin{center}
\caption{Quark Model Parameters.\label{modelparameters}}
\begin{tabular}{cccc}
\hline\noalign{\smallskip}
Quark masses   &$m_u=m_d$(MeV)     &313  \\
               &$m_s$(MeV)         &536  \\
               &$m_c$(MeV)         &1728 \\
               &$m_b$(MeV)         &5112 \\
\hline
Goldstone bosons   &$m_{\pi}(fm^{-1})$     &0.70  \\
                   &$m_{\sigma}(fm^{-1})$     &3.42  \\
                   &$m_{\eta}(fm^{-1})$     &2.77  \\
                   &$m_{K}(fm^{-1})$     &2.51  \\
                   &$\Lambda_{\pi}=\Lambda_{\sigma}(fm^{-1})$     &4.2  \\
                   &$\Lambda_{\eta}=\Lambda_{K}(fm^{-1})$     &5.2  \\
                   &$g_{ch}^2/(4\pi)$                &0.54  \\
                   &$\theta_p(^\circ)$                &-15 \\
\hline
Confinement             &$a_c$(MeV)     &101 \\
                   &$\Delta$(MeV)       &-78.3 \\
\hline
OGE                 & $\alpha_0$        &3.67 \\
                   &$\Lambda_0(fm^{-1})$ &0.033 \\
                  &$\mu_0$(MeV)    &36.976 \\
                   &$\hat{r}_0$(MeV)    &28.17 \\
\hline
\end{tabular} 
\end{center}
\end{table}

\begin{table}[htb]
\begin{center}
\caption{Meson Spectrum (unit: MeV)}
{\begin{tabular}{ccccccc}
\hline\noalign{\smallskip}
Meson   &Energy  &Experimental value \\
\hline\noalign{\smallskip}
$B_s^{0}$   &5368  &5366  \\
$\pi$       &139   &139 \\
$B_s^{*}$   &5410  &5415 \\
$\rho$      &772   &770 \\
$B^+$       &5281  &5279\\
$\bar{K}^0$ &494   &497 \\
$B^{*+}$    &5320  &5325\\
$\bar{K^*}^0$ &914 &892\\
$D_s^{-}$    &1953  &1968\\
$\bar{D}^0$  &1862 &1864 \\
\hline
\end{tabular} \label{meson spectrum}}
\end{center}
\end{table}

\section{Numerical Results} \label{Numerical Results}

In the present calculation, two structures of four-quark states, diquark-antidiquark
and meson-meson, are investigated. And in each structure, all possible states are considered.
For diquark-antidiquark structure, two color configurations, color antitriplet-triplet
($\bar{3}\times 3$) and sextet-antisextet ($6\times \bar{6}$) are taken into account.
For meson-meson structure, two color configurations, color singlet-singlet ($1\times 1$) and
octet-octet ($8\times 8$) are employed.

The calculation with the ordinary flavor symmetry, $SU(3)$ is first performed, i.e., no
Goldstone boson exchanges between $u,d,s$ and $b$ quark. In this case, the antisymmetrization
operator used is
\begin{equation}
{\cal A}=\sqrt{\frac{1}{2}}(1-P_{13})
\end{equation}
The results in this case are listed in Table \ref{result1}.
\begin{table}[htb]
\renewcommand\arraystretch{1.2} 
\caption{The energies of tetraquark system $su\bar{d}\bar{b}$ with flavor $SU(3)$ symmetry.$E_{th}^{theo}$ is the
theoretical threshold value and $E_{th}^{exp}$ represents the experimental threshold value.(unit: MeV)}
{\begin{tabular}{@{}ccccccc@{}}
\hline\noalign{\smallskip}
$qq-\bar{q}\bar{q}$   & $E_{\bar{3}\otimes 3}$ & $E_{6\otimes \bar{6}}$  &$E_{cc}$   & $E_{th}^{theo}$   & $E_{th}^{exp}$ \\
\hline
$su$$\bar{d}\bar{b}$  & 6406.0        & 6473.6            & 6360.0     & -                 & - \\
\hline
$q\bar{q}-q\bar{q}$         & $E_{1\otimes 1}$ & $E_{8\otimes 8}$  & $E_{cc}$    &    &   \\ \hline
$B_s^0 \pi$         & 5509.5           & 6443.5            & 5509.5     & 5507   & 5505 \\
$B_s^* \rho$       & 6185.5           & 6345.3            & 6185.5     & 6182   & 6185 \\
$B_s^0 \pi$-$B_s^* \rho$   & 5509.5   & 6324.3            & 5509.5     & 5507   & 5505  \\
$B^+ \bar{K}^0$    & 5776.8           & 6519.5            & 5776.8     & 5774   & 5776 \\
$B^{*+} \bar{K^*}^0$  & 6235.2        & 6403.9            & 6235.2     & 6233   & 6217 \\
$B^+ \bar{K}^0$-$B^{*+} \bar{K^*}^0$  & 5776.8 & 6376.9   & 5776.8     & 5774   & 5776 \\
\hline
\end{tabular} \label{result1}}
\end{table}

From the Table \ref{result1}, we can see that the two configurations of diquark-antidiquark structure,
$\bar{3}\times 3 $ and $6\times \bar{6}$, have similar energies, and the coupling between the two
configuration is rather strong. Nevertheless, the energy for diquark-antidiquark structure is too large
to be a natural candidate of the state $X(5568)$ in our calculation, although it could be a resonance
because of its color structure. With regard to meson-meson structure, the calculated energies approach
to the theoretical thresholds in all case. so no molecular structure formed in our model calculation.
In our calculations, the color singlet-singlet configurations always have the lower energies than that
of color octet-octet ones. The coupling between the two configurations is very small. The reason for
small coupling can be understood as follows. The effect of $K$-meson exchange is too weak to push the
energy of color singlet-singlet below the threshold, so the two colorless clusters tend to stay apart.
While two colorful clusters prefer stay close, the overlap between two configurations is small,
so the coupling from the exchange term of $K$-meson is small.
\begin{table}[htb]
\renewcommand\arraystretch{1.2} 
\caption{The energies of tetraquark system $su\bar{d}\bar{b}$ with flavor $SU(4)$ symmetry.$E_{th}^{theo}$ is the
theoretical threshold value and $E_{th}^{exp}$ represents the experimental threshold value. (unit: MeV)}
{\begin{tabular}{@{}ccccccc@{}}
\hline\noalign{\smallskip}
$qq-\bar{q}\bar{q}$   & $E_{\bar{3}\otimes 3}$ & $E_{6\otimes \bar{6}}$    & $E_{cc}$  & $E_{th}^{theo}$   & $E_{th}^{exp}$ \\
\hline
$su\bar{d}\bar{b}$   & 6397.6            & 6466.4           & 6351.0      & -                          & - \\
\hline
$q\bar{q}-q\bar{q}$         & $R_{1\otimes 1}$ & $E_{8\otimes 8}$  & $E_{cc}$    &    &   \\ \hline
$B_s^0 \pi$                    &5522.0            &6431.1            &5522.0     &5518   &5505 \\
$B_s^* \rho$                   &6282.7            &6324.3            &6182.5     &6177   &6185 \\
$B_s^0 \pi$-$B_s^* \rho$   &5522.0            &6306.1            &5521.0     &5518   &5505  \\
$B^+ \bar{K}^0$                &5717.6            &6440.1            &5717.6     &5715   &5776 \\
$B^{*+} \bar{K^*}^0$           &6204.6            &6277.2            &6204.5     &6202   &6217 \\
$B^+ \bar{K}^0$-$B^{*+} \bar{K^*}^0$  &5717.6 &6245.1            &5717.0     &5715   &5776 \\
\hline
\end{tabular} \label{result2}}
\end{table}

In the study of $N^{*}$ with hidden charm, the flavor $SU(4)$ symmetry plays an important role~\cite{Nstar}.
To see the effect of flavor $SU(4)$ symmetry, we extend our calculation from flavor $SU(3)$ symmetry to
$SU(4)$. In this case, the Goldstone boson exchanges including $\pi, K, \eta, B, B_s, \eta_b$,
totally fifteen pseudo-scalar mesons. For scalar mesons, we use effective $\sigma$-meson exchange instead of
sixteen scalar mesons~\cite{effective}.
 The mass of effective $\sigma$-meson takes the average of the quark
pairs, $u\bar{u}$, $d\bar{d}$, $s\bar{s}$ and $b\bar{b}$, due to its nature of flavor singlet of $SU(4)$.
In this work,we take different $m_{\sigma}^{eff}$ between two different quarks. For example,for $u$ and $s$ quark,
 $m_{\sigma}^{eff}=(2m_u+2m_s)/2$=849 MeV, or 4.3 fm$^{-1}$, the corresponding cutoff takes value
6.3 fm$^{-1}$. The results with flavor $SU(4)$ symmetry are shown in Table \ref{result2}. From the table,
we can see that the results are almost the same results as that of $SU(3)$. That is, no molecular state formed
and the energy for diquark-antidiquark structure is too large to be a candidate of the state $X(5568)$.
So in our quark model approach, the $X(5568)$ can not be explained as molecule or diquark-antidiquark state
under the constraint that the model describes the meson spectrum well.
Our results are different from the results of the previous work, e.g., Agaev's diquark-antiqdiquark explanation or
molecule of Ref.\cite{quarkmodel2}, but they support the analysis of Burns and Swanson~\cite{Burns}.
Because the state $X(5568)$ involves pseudo-scalar mesons, we think the main reason for the negative results
is that the Goldstone nature of the light pseudo-scalar mesons, which they have extraordinary small masses.

The calculation is also extended to the system composed of four different quarks: $s,u,\bar{c},\bar{d}$,
replacing the mass of heavy quark $\bar{b}$ by $\bar{c}$. The results are shown in Table \ref{resultudsc}.
\begin{table}[htb]
\renewcommand\arraystretch{1.2} 
\caption{The energies of tetraquark system $su\bar{c}\bar{d}$.$E_{th}^{theo}$ is the
theoretical threshold value and $E_{th}^{exp}$ represents the experimental threshold value. (unit: MeV)}
{\begin{tabular}{ccccccc}
\hline
 \multicolumn{6}{c}{$SU(3)$} \\ \hline
 $qq-\bar{q}\bar{q}$  & $E_{\bar{3} \otimes 3}$ & $E_{6 \otimes \bar{6}}$    & $E_{cc}$   & $E_{th}^{theo}$   & $E_{th}^{exp}$ \\
  \hline
$su\bar{c}\bar{d}$   &3059.0           &3073.9             &2983       &-                           &-\\
\hline
$SU(3)$ & $q\bar{q}-q\bar{q}$          & $E_{1\otimes 1}$ & $E_{8\otimes 8}$  & $E_{cc}$   &    &  \\ \hline
$D_s^- \pi$           &2095.1            &3080.6            &2095.1    &2092  &2107 \\
$\bar{K}^0 \bar{D}^0$ &2358.7            &3133.8            &2358.7    &2355   &2361 \\
\hline
 \multicolumn{6}{c}{$SU(4)$} \\ \hline
$qq-\bar{q}\bar{q}$   &$E_{\bar{3}\otimes 3}$ & $E_{6\otimes \bar{6}}$    & $E_{cc}$   & $E_{th}^{theo}$   & $E_{th}^{exp}$\\
\hline
$su\bar{c}\bar{d}$   &3023.4           &3073.9             &2943       &-                           &- \\
\cline{2-7}
$SU(4)$ &$q\bar{q}-q\bar{q}$     & $E_{1\otimes 1}$ & $E_{8\otimes 8}$  & $E_{cc}$   &    &  \\ \hline
 $D_s^- \pi$   &2088.7             &3043.6            &2088.6      &2085                        &2107 \\
 $\bar{K}^0 \bar{D}^0$  &2279.2    &3073.2            &2279.2    &2276                        &2361 \\
\hline
\end{tabular} \label{resultudsc}}
\end{table}

From the Table \ref{resultudsc}, we can obtain the same conclusion as that of $su\bar{c}\bar{d}$ system.
The masses of the system in the diquark-antidiquark structure are too large and in meson-meson
molecular structure approach the thresholds. Our calculation disfavor the existence of exotic state
$su\bar{c}\bar{d}$. The results are consistent with the general expectation that the heavier the
system, the stronger the states be bound.

\section{Summary} \label{Conclusion}
In this paper we have studied the new exotic resonance state $ X(5568)$ with the quantum numbers $J^P=0^+$,
which was observed recently by D0 Collaboration utilizing the collected data of $p\bar{p}$ collision.
The chiral quark model, which describes the meson spectrum well, is employed to do the calculation.
Two structures: diquark-antidiquark and meson-meson, with flavor symmetries, $SU(3)$ and $SU(4)$,
are investigated. We find that the masses of $us\bar{b}\bar{d}$ with diquark-antidiquark structure
are too high to be candidate of the state $X(5568)$ and no molecular structure can be formed.
The calculation is extended to $us\bar{c}\bar{d}$ system, the same conclusion is obtained.

Because of the quark contents of the system, the pseudo-scalar mesons are involved. The extraordinary
small masses of these Goldstone bosons disfavors the existence of the exotic. Our results agree with
the analysis of Burns and Swanson. The recent preliminary results of LHCb collaboration do not confirm
the state $X(5568)$, So more experimental and theoretical work are needed to clarify the situation.

\begin{acknowledgement}
The work is supported partly by the National Natural Science Foundation of China under Grant
Nos. 11175088 and 11535005.
\end{acknowledgement}

\end{document}